# Cavity enhanced storage - preparing for high efficiency quantum memories

M Sabooni, S Tornibue Kometa, A Thuresson, S Kröll, L Rippe

Department of Physics, Lund University, P.O. Box 118, SE-22100 Lund, Sweden

E-mail: `mahmood.sabooni@fysik.lth.se`

**Abstract.** Cavity assisted quantum memory storage has been proposed [PRA 82, 022310 (2010), PRA 82, 022311 (2010)] for creating efficient (close to unity) quantum memories using weakly absorbing materials. Using this approach we experimentally demonstrate a significant ($\sim$ 20-fold) enhancement in quantum memory efficiency compared to the no cavity case. A strong dispersion originating from absorption engineering inside the cavity was observed, which directly affect the cavity line-width. A more than 3 orders of magnitude reduction of cavity mode spacing and cavity line-width from GHz to MHz was observed. We are not aware of any previous observation of several orders of magnitudes cavity mode spacing and cavity line-width reduction due to slow light effects.

## 1. Introduction

Coherent and reversible mapping of a quantum state between light and matter, as a flying and stationary qubit, is an essential step towards scalability of quantum information processing [1, 2]. Such an optical quantum memory would be important for storing and synchronizing the output quantum states of several quantum gates in quantum computation protocols [3]. In addition, quantum memories may also play a crucial role in enabling long distance ($>$ 150 $km$) quantum communication [4], and in other applications [5]. Most ensemble based optical quantum memory [5, 6] protocols belongs to one of two main groups. One where the protocols are based on electromagnetically induced transparency (EIT) [7, 8, 9] and one where they are based on the coherent rephasing (echo) effect [10] such as controlled reversible inhomogeneous broadening (CRIB) [11, 12], atomic frequency comb (AFC) [13, 14, 15, 16, 17, 18], gradient echo memory (GEM) [19, 20, 21], and Revival of silenced echo (ROSE) [22].

There are specific criteria for assessing quantum memory performance based on the application in mind but generally we could list the following key parameters: efficiency, fidelity, storage time, multi-mode storage capacity, and operational wavelength [5]. The scope of this paper is mainly to demonstrate how to improve the memory efficiency of weakly absorbing materials. The memory efficiency is here defined as the energy of the pulse recalled from the memory divided by the energy of the pulse sent in for storage in the memory. At the quantum level (weak coherent pulses), storage efficiency basically corresponds to the probability that the stored information is retrieved. Fidelity is simply defined as the overlap of the quantum state wave-function of the recalled photon with that of the one originally sent in for storage. In the optimum fidelity case, the wave-function for the input and output photons are identical. An alternative definition is conditional fidelity, i.e. the overlap conditional on that the photon was emitted by the memory [5]. Based on the application, there will also be requirements on longer storage time [7, 23, 24, 25] and on-demand retrieval. The capacity to store several photons (modes) at the same time in the memory is called multi-mode storage capability. Multi-mode storage is crucial in probabilistic applications to boost the occurrence rate of events [26, 16, 27] and the AFC concept has been shown to be particularly suitable for multi-mode applications [28].

In the AFC protocol, which is employed in this paper, the combination of photon-echo and spin-wave storage allows on-demand storage. However, the present experiment does not include spin storage. The principle of the AFC protocol preparation is shown in Fig. 1. A sequence of narrow peaks, equidistance in frequency, with a peak optical absorption depth $d$ and background absorption $d_0$ is created by optical pumping and/or targeted state-to-state transfer pulses. The input field for storage, with a spectral distribution larger than the peak separation $\Delta$ in Fig. 1 and narrower than the AFC bandwidth could be totally absorbed by a sequence of narrow peaks. Afterwards, the input field will excite all the atoms in the AFC structure coherently. The collective coherent state distributed over AFC peaks separated by $\Delta$ will rapidly dephase into a non-collective state. Well separated equidistance AFC peaks will rephase and lead to a collective emission after a time $1/\Delta$.

For high efficiency quantum state storage one basic requirement is to have enough absorption depth to fully transfer all the input pulse energy to the atomic media. However, in strongly absorbing media one also needs to eliminate the reabsorption of the emitted echo in the medium. This can be done by shifting the absorber using the linear Stark effect with an electric field gradient where the output field will be emitted in the backward [11] or forward [20] direction. On the other hand, the strong absorption requirement clearly will pose some limitations on the material used. Thus it would be very important to enhance the absorption of materials with long coherence time but low absorption (e. g. Europium).

Interestingly, complete absorption of the input pulse can also be obtained for weakly absorbing materials (that for example could have longer coherence time, such as $Eu^{3+}:Y_2SiO_5$ [31]) if they are inserted in a cavity [32, 33]. This cavity should satisfy the impedance matched condition for achieving the complete absorption of the input pulse. To do this, the absorption during one round trip of the cavity should be exactly equal to the transmission of the input coupling mirror and also there should be no other losses during the cavity round trip. Obtaining this (impedance-matched) condition is a vital step towards efficient quantum memories using weakly absorbing media. In this work, we employed a cavity configuration around our absorber material and demonstrate more than one order of magnitude improvement in memory efficiency compared to that obtained without the cavity. The Praseodymium absorption line in $Pr^{3+}:Y_2SiO_5$ was tailored based on the AFC protocol [13] and an impedance matched cavity configuration [32] is employed to mainly enhance the quantum memory efficiency. On-demand recall as well as longer quantum memory storage time could be achievable via spin-wave storage [14, 23] but this is beyond the scope of the prepared paper. Cavity enhanced spin wave to photon conversion efficiency results have for example been demonstrated n atomic gases [34, 35].

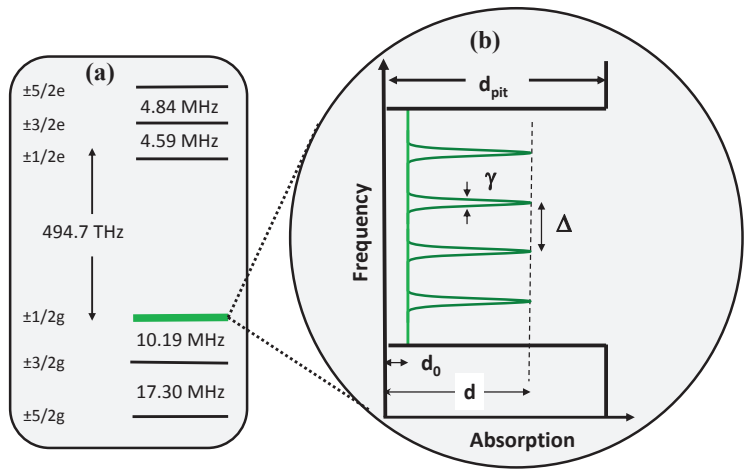

**Figure 1.** (Color online) (a) The hyperfine splitting of $^3H_4 - {}^1D_2$ transition of the site I $Pr^{3+}:Y_2SiO_5$ is shown [29, 30]. (b) The $Pr$ absorption engineering based on atomic frequency comb (AFC) protocol. $d_{pit}$ is the absorption depth outside of the AFC structure spectral region which is situated in what is called a spectral pit, $d$ is the peak optical depth while $d_0$ is background absorption. $\Delta$ is the AFC peak separation and $\gamma$ is the full-width at half-maximum (FWHM) of the AFC peaks. The finesse of the AFC structure will be $F_{AFC} = \frac{\Delta}{\gamma}$. The effective absorption of the AFC structure is defined as $\tilde{d} = \frac{d-d_0}{F_{AFC}}$.

The paper is organized in the following way. In Sec. 2 we give an overview of the experimental setup in particular what regards the cavity design. In Sec. 3 the effect of having an absorbing quantum memory material inside the cavity is discussed. In Sec. 4 we show quantum memory efficiency results following the preparation steps towards the AFC memory demonstration, and in Sec. 5 a further discussion about the strong dispersion and cavity free spectral range effect is included. Conclusions are given in Sec. 6.

## 2. Experimental setup

This section discusses the experimental setup and is divided into two main parts. The first part describe the stabilized laser source, and the modulator setup which generates the pulses, while the second part outlines the cavity crystal design as shown in Fig. 2a. A 6 $Watt\ Nd:YVO_4$ coherent Verdi-V6 laser at 532 nm pumps a Coherent 699-21 dye laser which generates $\sim 500\ mW$ laser radiation at 605.977 nm. This wavelength matches the $^3H_4 - {}^1D_2$ transition of praseodymium ions doped into yttrium silicate ($Pr^{3+}:Y_2SiO_5$) [29, 30].

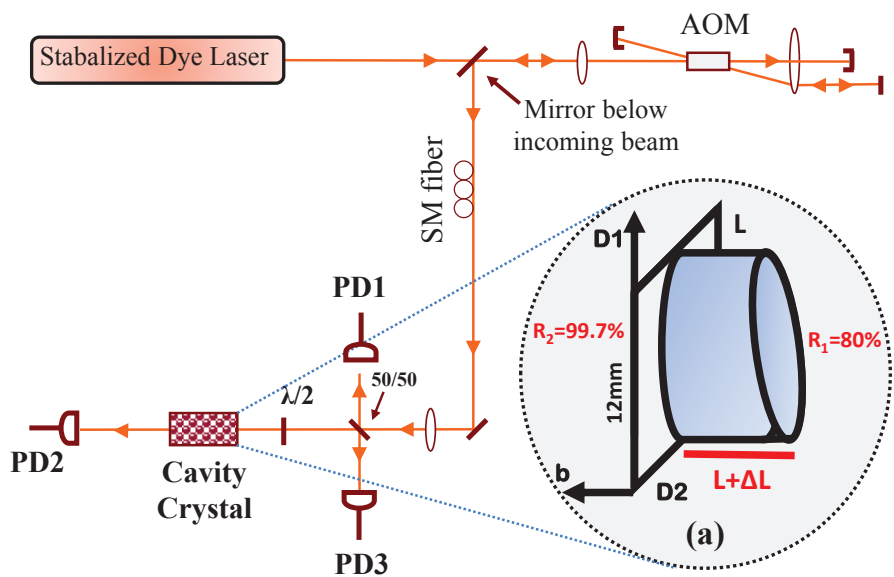

**Figure 2.** (Color online) Experimental set up. A double pass AOM is employed to tailor the optical pulses which comes from a narrow (kHz) line-width laser. Photo diode (PD1) monitor the input light to the cavity. Two other photo diodes monitor the transmitted (PD2) and reflected (PD3) light from the cavity. The cavity crystal was designed to have small wedge, such that the length L, differers by about $\approx > \Delta L = 200\ nm$. In practice, the wedged surface is not completely flat, see Fig. 3. (a) The input and output facet of the crystal have 80% and 99.7% reflectivity, respectively.

The laser stabilization system which uses hole-burning in a $Pr^{3+}:Y_2SiO_5$ crystal in a separate cryostat, as frequency reference, improves the frequency stability of the laser to $\approx 1\ kHz$ line-width [36]. Good frequency stability is crucial for coherent interaction experiments such as the ones described here. The $^3H_4 - {}^1D_2$ transition in $Pr^{3+}:Y_2SiO_5$ has an excited state coherence time ($T_2$) of 152 $\mu s$ [29]. A double pass acousto-optic modulator (AA. opto-electronics St.200/B100/A0.5-VIS, AOM) with 200 $MHz$ center frequency is employed for tailoring the laser light amplitude, phase and frequency and carry out the required pulse shaping in the experiment. An extensive MATLAB code calculates the pulse shapes and a 1-GS/s arbitrary waveform generator (Tektronix

AWG520), generates the RF signal which drives the AOM that modulates the light. The spatial laser mode is cleaned by passing light through a single-mode (SM) fiber and then a 50/50 beam splitter is used to obtain the reference signal at the photo diode (PD1) before the cavity crystal. The transmitted light through the cavity will be detected at PD2 in Fig. 2. The reflected light from the cavity will be detected by the third photo diode (PD3) in Fig. 2. The PD1 and PD3 are PDB150A photo detectors from Thorlabs while PD2 is a Hamamatsu S5973 photo diode which is connected directly to a Femto DHP CA-100 transimpedance amplifier. The beam diameter at the cavity crystal position is set to be about $80\,\mu m$. The crystal was immersed in liquid helium at 2.1 K to ensure the optical coherence time ($T_2$) of the praseodymium is not shortened by phonon interaction. A $\lambda/2$ plate was used before the crystal to control the light polarization direction with respect to the crystal axes and the $Pr$ transition dipole moment direction [37].

The cavity crystal is made out of 0.05 % praseodymium ions doped into yttrium silicate ($Pr^{3+}:Y_2SiO_5$). It has a 12 $mm$ diameter and is (except for the small wedge) cut along the $D_1 \times D_2$ plane with a thickness of 2 $mm$ as shown in Fig. 2a. The cavity mirrors are created by coating the 12 $mm$ diameter crystal surfaces. Multiple layer stacks of $SiO_2$ and $ZrO_2$ are designed and applied for these coating surfaces by Optida. The light propagation direction is selected to be along the crystal b axis as shown in Fig. 2. The input and output cavity facets are designed to have reflectivities of $R_1 = 80\,\%$ and $R_2 = 99.7\,\%$, respectively. The losses because of the $Pr$ ion absorption through the double pass of the cavity should be adjusted to compensate the unequal cavity surface coating to achieve the impedance matched cavity condition [32, 33]. The impedance matched cavity condition occurs when the directly reflected field and a transmitted field coming from the field circulating inside the cavity cancel each other at the input surface and consequently the reflected intensity vanishes.

## 3. Cavity without and with absorber

The calculated cold cavity (i.e. neglecting any $Pr$ absorption effects) transmission frequency spacing ($\frac{c}{2nL}$) for this crystal with $n = 1.8$ and $L = 2\ mm$ is 41.6 $GHz$. The theoretical finesse of this cavity based on the mentioned mirror reflectivities is $\approx 25$ which corresponds to a spectral line-width of 1.5 $GHz$. Scanning the laser light across the cold cavity (off-resonance with the Pr ion absorption line) shows a cavity transmission line-width of about 2.5 $GHz$ which corresponds to a finesse of about $\approx 16$. Part of the reduction could be related to the wedge design causing a small beam walk off as the light bounces between the mirrors.

To spectrally match the cavity transmission profile to the inhomogeneous absorption line of $Pr$ [29], the crystal was designed to have a small wedge as shown in Fig. 2a, such that the length L, differs by about 200 $nm$. In this way the cavity transmission spectrum could be translated well over one free spectral range. To accurately match the cavity transmission peak to the $Pr$ absorption line, a translation stage with sub $\mu$m accuracy (Attocube system, ANCz150) is employed to translate the crystal perpendicular to the input beam.

The parallelism of the cavity surfaces is determined by an interference experiment. A collimated coherent beam ($\lambda = 606\ nm$) with diameter $> 12\ mm$, impinged almost perpendicularly to the crystal surface and the reflected pattern was monitored. Fig. 3 shows the non-uniform interference pattern before coating the cavity surface. One could estimate the cavity parallelism in Fig. 3 as the changes in the cavity length is $\lambda_{material}/2$ between two bright or dark fringes in the cavity. Since Fig. 3 shows the light reflected from the cavity, the dark fringes are the areas of larger transmission.

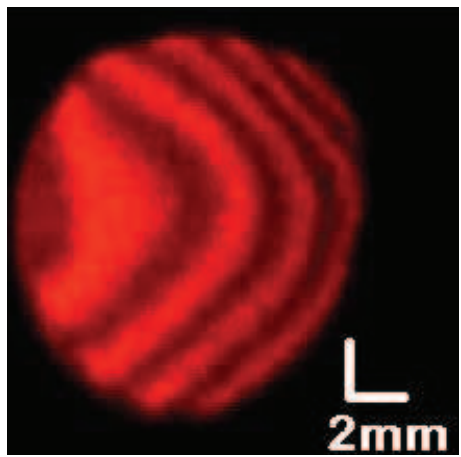

**Figure 3.** (Color online) An interference experiment is used to investigate the parallelism of the cavity surfaces. A collimated laser beam at $\lambda = 606nm$ was impinging perpendicularly to the 12 $mm$ diameter cavity crystal surface. The figure shows the reflected light from the cavity surface displayed on a screen.

For quantum state storage, the AFC protocol [13] is chosen for engineering the absorption profile of the inhomogeneously broadened $Pr$ ion. The heart of the AFC protocol is using a spectrally periodic medium for enhancing the storage capacity. Using a frequency domain interpretation, the periodic structure shown in Fig.1b can be viewed as a spectral grating. The storage pulse tuned to the AFC structure is diffracted in time and produces an echo [38]. Based on a time domain interpretation, this will be interpreted as a rephasing of a set of dipoles in time [13]. The storage pulse will be absorbed by AFC peaks well separated in frequency ($\Delta$ in Fig.1b). The collective emission of AFC peaks will be in phase after time $1/\Delta$ generating an echo of the input pulse.

The first step towards this preparation is to make a zero absorption window within the inhomogeneous $Pr$ ion absorption profile. A series of optical pumping pulses is employed to create an 18 $MHz$ transmission window (spectral pit) within the $\approx 9\ GHz$ inhomogeneous profile of $Pr^{3+}:Y_2SiO_5$. A detailed discussion about what pulses can be used to create the transmission window is presented in Ref. [17]. This transmission window is called spectral pit in Ref.[30] which we will use afterwards.

It turns out that a spectral transmission window (spectral pit) might change the cavity mode spacing by several orders of magnitude due to slow light effects. This will now be briefly explained by considering a dispersive media inside a cavity as follows (see Eq. 11.58 p.437 Ref. [39]):

$$\Delta\nu_{mode} = \nu_{q+1} - \nu_q = \frac{1}{2L}\frac{c_0}{n_g(\nu)} = \frac{v_g}{2L} \quad (1)$$

$$n_g(\nu) = \frac{c_0}{v_g} = \frac{\partial(\nu n_r(\nu))}{\partial \nu} = n_r(\nu) + \nu\frac{dn_r(\nu)}{d\nu} \quad (2)$$

Where q is an integer, L is the cavity length, $c_0$ is the speed of light in vacuum, $n_g(\nu)$ is the refractive index for the group velocity, $n_r(\nu)$ is the real part of the refractive index for the phase velocity and $v_g$ is the group velocity. It is important to note that the index of refraction in Eq. 1 is not the index of refraction for the phase velocity, $n$, but the index of refraction of the group velocity, $n_g$. This is sometimes overlooked, but in a low dispersion medium the two indices of refraction are the same ($n \approx n_g$) so often the dispersion is not important. The real and imaginary parts of the susceptibility are connected via the Kramers-Kronig relation as follows:

$$Re(\chi(\nu)) = \frac{2}{\pi}\int_0^\infty \frac{\nu' Im(\chi(\nu'))}{\nu'^2 - \nu^2} d\nu' \quad (3)$$

$$Im(\chi(\nu)) = \frac{2}{\pi}\int_0^\infty \frac{\nu' Re(\chi(\nu'))}{\nu'^2 - \nu^2} d\nu' \quad (4)$$

Therefore by measuring the absorption spectrum $\alpha(\nu)$ which is directly related to $Im(\chi(\nu))$ and use this as an input to Eq. 3, one can calculate $Re(\chi(\nu))$ which is proportional to $n_r(\nu)$ and then calculate the dispersion ($\frac{dn_r(\nu)}{d\nu}$). In the cold cavity case where there is no absorption and consequently also no dispersion, the $n_r(\nu) \gg \nu\frac{dn_r(\nu)}{d\nu}$ regime hold while in the case of a spectral pit we have $n_r(\nu) \ll \nu\frac{dn_r(\nu)}{d\nu}$. Actually, using the experimentally recorded absorption $\alpha(\nu)$ as input to Eq. 3 gives a dispersion term ($\nu\frac{dn_r(\nu)}{d\nu}$) in Eq. 2 which is 3-4 orders of magnitude larger than $n_r(\nu)$. This means that the mode spacing for frequencies inside the spectral transmission window now only is of the order of ten *MHz*.

To experimentally monitor the created transmission window a weak minimally-disturbing pulse was frequency chirped at a rate of 1 *MHz/µs* across the corresponding frequency region. In the absence of a cavity this procedure typically displays an $\approx$ 18 *MHz* spectral pit (see e.g. Fig. 1 in Ref. [40]). For the cavity case, only a much more narrow transmission line ($\approx$2 *MHz*) inside the spectral region was observed. This sharp transmission lines turns out to be about 3 orders of magnitude narrower than the cold cavity transmission peaks. The cavity finesse ($F_{cav}$) is primarily determined by the mirror reflectivity. Therefore $F_{cav}$ should be independent of the spectral pit (transmission window) properties provided that possible residual absorption in the spectral pit is small. Based on the assumption of a constant cavity finesse, the ratio between the cavity free spectral range ($\Delta\nu$) and $\delta$ will be constant. Thereby, it follows from Eq. 1 that the cavity transmission line-width is proportional to the group velocity ($v_g$). Thus the 2 $MHz$ transmission peak is largely consistent with obtained results based on the theoretical calculation above. Two cavity transmission peaks that are translated by changing the cavity length are shown in Fig. 4. The cavity transmission peaks was translated by moving the slightly wedged crystal perpendicular to the beam propagation direction. The reduction of the cavity transmission line-width is also due to the strong dispersion which is created by the spectral pit inside the cavity.

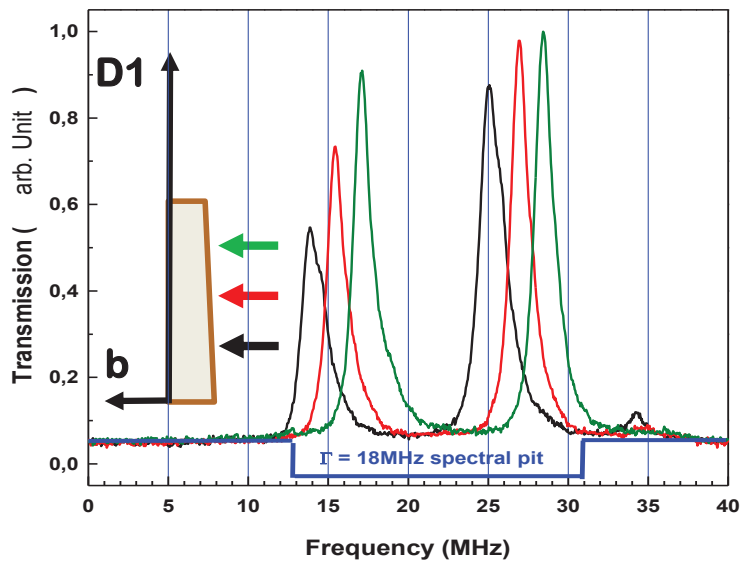

**Figure 4.** (Color online) Three cavity transmission spectra taken at different positions through the crystal as indicated by the arrows are shown. The cavity crystal with about 2 $mm$ length and 80% and 99.7% mirror reflectivity while the $Pr$ absorption line was tailored to create $\approx$ 18 $MHz$ spectral pit within the cavity. The spectral pit is responsible for the 3 orders of magnitude cavity mode reduction from GHz to MHz. The maximum difference in cavity transmission spectrum as the beam was translated at the input surface is about 3 $MHz$, which corresponds to about 40 $nm$ cavity length changes. The designed wedge on the crystal surface gives us enough freedom to translate the cavity transmission between two cavity modes which is $\approx$ 11 $MHz$ here.

In fact although there are experimental observation of changes in the cavity mode spacing in active media of more than an order of magnitude [41, 42], we are not aware of any effects as large as the 3-4 orders of magnitude observed here. Details of this effect are however beyond the scope of the present paper and will be addressed in a separate publication [43].

## 4. AFC preparation and memory efficiency

Each AFC peak shown in Fig. 1b is prepared within the created transmission window (spectral pit) by employing two consecutive complex hyperbolic secant [30] pulses to optically pump $Pr$ ions from $|\pm 5/2g\rangle \to |\pm 5/2e\rangle$ and $|\pm 5/2e\rangle \to |\pm 1/2g\rangle$ (see Fig.

1a). This procedure is carried out four times, each time with a new center frequency shifted by $\Delta$ to create an AFC structure as shown in Fig. 1b. In this way the AFC structure is engraved in the $|\pm 1/2g\rangle$ state. To be able to use the cavity configuration and improve the storage efficiency, the AFC structure needs to be engraved within the cavity transmission window. Therefore, the cavity transmission line-width ($\delta$) needs to be optimized to cover the whole AFC structure bandwidth. As discussed in Ref.[40] the group velocity can be written: $v_g = \frac{2\pi\Gamma}{\alpha}$ where $\Gamma, \alpha$ are spectral pit width and the absorption coefficient outside the spectral pit, respectively. It was therefore concluded that in $0.05\% : Pr:YSO$, where the maximum spectral pit window width ($\Gamma$) is limited to $18\,MHz$ and at the line center absorption coefficient ($\alpha$) is about $2000\,m^{-1}$, the only possibility for the cavity transmission line of width $\delta$ to cover the AFC bandwidth was to obtain a smaller $\alpha$ by moving the whole spectral pit to a part of the inhomogeneous profile where the absorption was lower than at the inhomogeneous line center. The created spectral pit and the AFC peaks will affect the dispersion and finally the cavity mode structure (see Eq. 1). Therefore, the shape of the spectrum is a combination of the effect from the AFC absorption pattern and the cavity dispersion effect.

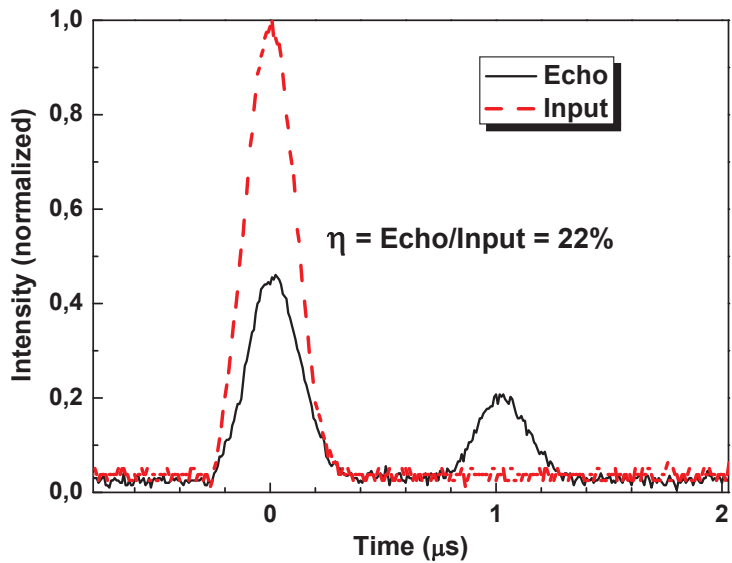

**Figure 5.** (Color online) Storage efficiency measurement. A 250 ns FWHM Gaussian pulse tuned to an AFC structure with 4 peaks separated by $\Delta = 1\,MHz$. 22 % of the input energy (red dashed line) was retrieved in the echo (black solid line) after $1\,\mu s$.

The next step after the AFC preparation is to send the storage pulse tuned to the AFC structure center frequency through mirror $R_1$ (see Fig. 2). For this purpose a Gaussian pulse with $\tau_{FWHM} = 250\,ns$ was employed. For an AFC structure with peak separation $\Delta$, there will be a rephasing after a delay $1/\Delta$ [13]. The black solid trace in Fig. 5 is the signal when the storage pulse is tuned to the AFC structure center frequency and impinged onto mirror $R_1$. We define the memory efficiency as ratio of the output (echo) energy divided by the input pulse energy. To obtain a reference signal, for the input energy to the memory, the cavity was turned ∼ 180° such that the input Gaussian storage pulse impinged on the mirror with reflectivity $R_2$ (nearly 100 % reflection) and the reflection was detected at the reflection detector (PD3). This signal represent the input signal shown with the (red) dashed line in Fig. 5 and corresponds to the signal strength that would be observed if the memory efficiency would be 100%. Reference detector PD1 was employed to compensate the measurements for laser fluctuation from one measurement to another. 22 % of the input energy was retrieved in the echo after 1 $\mu$s as shown in Fig. 5.

## 5. Strong dispersion and cavity free spectral range

Finally, an interesting observation is that the requirement of matching the absorption with the mirror reflectance in combination with the slow light effect when creating spectral pit and the AFC structure, may put restrictions on the system performance as shown by the following calculation: According to Eq. 1 the mode spacing of a cavity with length L is given by $\Delta \nu = \frac{v_g}{2L}$ and from $v_g = \frac{2\pi \Gamma L}{d_{pit}}$ [40, 44] the cavity line-width, $\delta$, could be written:

$$\delta = \frac{\Delta \nu}{F_{cav}} = \frac{\pi \Gamma}{F_{cav} d_{pit}} \qquad (5)$$

where $F_{cav}$ is the cavity finesse. Therefore, as a general statement and regardless of any memory preparation protocol, one can achieve the ratio $\frac{\delta}{\Gamma} = \frac{\pi}{F_{cav} d_{pit}}$. To explore the cavity condition on the AFC preparation, one could write the cavity finesse

$$F_{cav} = \frac{\pi R^{1/4}}{1 - \sqrt{R}} \qquad (6)$$

where the mirror reflectance (R) should satisfy the impedance matching condition ($R = exp(-2\tilde{d})$) and $\tilde{d} = \frac{d}{F_{AFC}}$. $\tilde{d}$ is the effective absorption ($d_0 = 0$) according to Fig. 1. Considering $2\tilde{d} \ll 1$ we could simplify Eq. 6 as

$$F_{cav} \approx \frac{\pi}{\tilde{d}} \qquad (7)$$

Now we can rewrite the cavity line-width:

$$\delta = \Gamma \frac{\tilde{d}}{d_{pit}} = \frac{\Gamma}{F_{AFC}} \qquad (8)$$

Therefore, the cavity line-width for a matched cavity to first order just depends on the spectral pit width and the finesse of the AFC structure. This may cause complications because to minimise memory loss due to dephasing, $F_{AFC}$ should be as high as possible and $\Gamma$ is limited by the hyperfine splitting in the material and cannot easily be varied freely.

This can affect the fitting of the AFC structure under the cavity line-width since the cavity line-width might be not wide enough. But, practically we do not expect this limitation to become important until one gets into the high efficiency ($\eta > 90\%$) regime where it is important to suppress memory loss due to the dephasing factor ($exp(\frac{-7}{F_{AFC}})$)

*Cavity enhanced storage - preparing for high efficiency quantum memories*     11

in Eq. 9) by choosing higher AFC finesse ($F_{AFC} > 9$). Even in this case, the cavity linewidth will readily still be close to the exited state hyperfine transition separation which is the bandwidth limitation of the AFC protocol as discussed in Ref. [13]. On the other hand the AFC peaks themselves are very narrow absorption structures and consequently will create strong dispersion effects. The effects will modify the transmission structure and this may offer further possibilities, however, as previously stated, a detailed analysis is beyond the scope of the present paper.

To estimate the enhancement in storage efficiency attained by employing the cavity configuration we calculate what signal would have been expected if the end faces were un-coated. The AFC storage efficiency, $\eta$, is given by [45]

$$\eta = \tilde{d}^2 exp(-\tilde{d}) exp(\frac{-7}{F_{AFC}^2}) exp(-d_0) \qquad (9)$$

where $\tilde{d} = \frac{d-d_0}{F_{AFC}}$ is the effective absorption, $d_0$ is the background absorption, and $F_{AFC} = \frac{\Delta}{\gamma}$ as shown in Fig. 1b. This equation has been shown to agree very well with experiment in Refs. [15, 17]. According to the Eq. 5, we can estimate the ratio between the absorption depth of the pit at the spectral position where the quantum memory was prepared, $d_{pit}^{QM}$, and the absorption depth of the pit at the line center, $d_{pit}^c$, (See Fig. 1b) as following:

$$\frac{d_{pit}^{QM}}{d_{pit}^c} = \frac{\delta_c}{\delta_{QM}} \qquad (10)$$

where $\delta_{QM}$ and $\delta_c$ are the cavity line-width at the quantum memory preparation spectral position and the line center, respectively (with the assumption that $\Gamma$ and $F_{cav}$ are similar in these two cases). For the present setup, it turns out we have $\delta_c = 1.5\,MHz$, $\delta_{QM} = 8\,MHz$, and $d_{pit}^c = 4$ which gives $d_{pit}^{QM} = 0.75$. To our knowledge $d < d_{pit}^{QM}$ in all papers published so far, thus we might chose the actual optical depth of the AFC structure $d = 0.75$. From Eq. 9, considering negligible background absorption $d_0$, and $F_{AFC} = 7$ (which is our best guess for the present work) we get an efficiency of $< 1\%$. Thus the estimate is that we at least have observed a 20-fold enhancement using the cavity.

The main experimental limitations in the present setup for achieving unity storage and retrieval efficiency, which is predicted theoretically [32, 33], are the following. First, as shown by the black solid trace in Fig. 5 about 50% of the input pulse is reflected directly at time $t = 0$. Therefore, to improve the storage efficiency one needs to increase the input pulse absorption. Limited absorption means, then the impedance-matched condition is not fulfilled. This is partly related to poor spatial mode matching. However, it is also possible that the impedance-matching condition in our cavity crystal could be improved by locking the laser further from the $Pr$ line center ($> 5\,GHz$) where the absorption is lower. This could however not be fully tested in the present setup since it is difficult to select the laser frequency stabilization point far from the line center for the stabilization setup based on a spectral hole [36]. Second, after optimizing the absorption, improving the retrieval part of the memory maybe possible by improving the finesse of the prepared AFC structure ($F_{AFC}$).

## 6. Conclusion

In conclusion, we demonstrate a $\sim$ 20-fold enhancement in quantum memory efficiency for cavity assisted quantum memories compared to the no cavity case. This is a proof of principle for achieving high quantum memory efficiency using weakly absorbing media. In addition, we demonstrate $\sim$ 3 order decrease in cavity mode spacing using slow light effects in a dispersive medium. We are not aware of any previous observations of cavity mode spacing changes of this order of magnitude due to slow light effects.

## 7. Acknowledgments

This work was supported by the Swedish Research Council, the Knut & Alice Wallenberg Foundation, the Crafoord Foundation and the EC FP7 Contract No. 247743 (QuRep). We are grateful to Dr. Mikael Afzelius for several valuable discussions.

## 8. References

[1] Kimble H J 2008 *Nature* **453** 1023–1030
[2] Ritter S, Nolleke C, Hahn C, Reiserer A, Neuzner A, Uphoff M, Mucke M, Figueroa E, Bochmann J and Rempe G 2012 *Nature* **484** 195–201
[3] Nunn J, Langford N K, Kolthammer W S, Champion T F M, Sprague M R, Michelberger P S, Jin X M, England D G and Walmsley I A 2012 *arXiv:1208.1534v1*
[4] Duan L M, Lukin M D, Cirac J I and Zoller P 2001 *Nature* **414** 413–418
[5] Simon C, Afzelius M, Appel J, de la Giroday A B, Dewhurst S J, Gisin N, Hu C Y, Jelezko F, Kröll S, Muller J H, Nunn J, Polzik E S, Rarity J G, De Riedmatten H, Rosenfeld W, Shields A J, Skoeld N, Stevenson R M, Thew R, Walmsley I A, Weber M C, Weinfurter H, Wrachtrup J and Young R J 2010 *European Physical Journal D* **58** 1–22
[6] Tittel W, Afzelius M, Chanelière T, Cone R L, Kröll S, Moiseev S A and Sellars M 2010 *Laser & Photonics Reviews* **4** 244–267
[7] Longdell J J, Fraval E, Sellars M J and Manson N B 2005 *Physical Review Letters* **95** 063601
[8] Hetet G, Peng A, Johnsson M T, Hope J J and Lam P K 2008 *Physical Review A* **77** 012323
[9] Fleischhauer M, Imamoglu A and Marangos J P 2005 *Reviews of Modern Physics* **77** 633–673
[10] Moiseev S A and Kröll S 2001 *Physical Review Letters* **87** 173601
[11] Nilsson M and Kröll S 2005 *Optics Communications* **247** 393–403
[12] Kraus B, Tittel W, Gisin N, Nilsson M, Kröll S and Cirac J I 2006 *Physical Review A* **73** 020302
[13] Afzelius M, Simon C, de Riedmatten H and Gisin N 2009 *Physical Review A* **79** 052329
[14] Afzelius M, Usmani I, Amari A, Lauritzen B, Walther A, Simon C, Sangouard N, Minar J, de Riedmatten H, Gisin N and Kröll S 2010 *Physical Review Letters* **104** 040503
[15] Chanelière T, Ruggiero J, Bonarota M, Afzelius M and Le Gouet J L 2010 *New Journal of Physics* **12** 023025
[16] Bonarota M, Le Gouët J L and Chanelière T 2011 *New Journal of Physics* **13** 013013
[17] Amari A, Walther A, Sabooni M, Huang M, Kröll S, Afzelius M, Usmani I, Lauritzen B, Sangouard N, de Riedmatten H and Gisin N 2010 *Journal of Luminescence* **130** 1579–1585
[18] Sabooni M, Beaudoin F, Walther A, Lin N, Amari A, Huang M and Kröll S 2010 *Physical Review Letters* **105** 060501

[19] Hetet G, Longdell J J, Sellars M J, Lam P K and Buchler B C 2008 *Physical Review Letters* **101** 203601
[20] Hedges M P, Longdell J J, Li Y and Sellars M J 2010 *Nature* **465** 1052–1056
[21] Hosseini M, Sparkes B M, Campbell G, Lam P K and Buchler B C 2011 *Nature Communications* **2** 174
[22] Damon V, Bonarota M, Louchet-Chauvet A, Chanelière T and Le Gouët J L 2011 *New Journal of Physics* **13** 093031
[23] Heinze G 2013 *et. al. (to be published)*.
[24] Steger M, Saeedi K, Thewalt M L W, Morton J J L, Riemann H, Abrosimov N V, Becker P and Pohl H J 2012 *Science* **336** 1280–1283
[25] Maurer P C, Kucsko G, Latta C, Jiang L, Yao N Y, Bennett S D, Pastawski F, Hunger D, Chisholm N, Markham M, Twitchen D J, Cirac J I and Lukin M D 2012 *Science* **336** 1283–1286
[26] Simon C, de Riedmatten H, Afzelius M, Sangouard N, Zbinden H and Gisin N 2007 *Physical Review Letters* **98** 190503
[27] Usmani I, Afzelius M, de Riedmatten H and Gisin N 2010 *Nature Communications* **1** 12
[28] Nunn J, Reim K, Lee K C, Lorenz V O, Sussman B J, Walmsley I A and Jaksch D 2008 *Physical Review Letters* **101** 260502
[29] Equall R W, Cone R L and Macfarlane R M 1995 *Physical Review B* **52** 3963–3969
[30] Rippe L, Nilsson M, Kröll S, Klieber R and Suter D 2005 *Physical Review A* **71** 062328
[31] Lauritzen B, Timoney N, Gisin N, Afzelius M, de Riedmatten H, Sun Y, Macfarlane R M and Cone R L 2012 *Physical Review B* **85** 115111
[32] Afzelius M and Simon C 2010 *Physical Review A* **82** 022310
[33] Moiseev S A, Andrianov S N and Gubaidullin F F 2010 *Physical Review A* **82** 022311
[34] Simon J, Tanji H, Thompson J K and Vuletic V 2007 *Physical Review Letters* **98** 183601
[35] Bao X H, Reingruber A, Dietrich P, Rui J, Dück A, Strassel T, Li L, Liu N L, Zhao B and Pan J W 2012 *Nature Physics* **8** 517–521
[36] Julsgaard B, Walther A, Kröll S and Rippe L 2007 *Optics Express* **15** 11444–11465
[37] Sun Y, Wang G M, Cone R L, Equall R W and Leask M J M 2000 *Physical Review B* **62** 15443–15451
[38] Sonajalg H and Saari P 1994 *Journal of the Optical Society of America B-optical Physics* **11** 372–379
[39] Siegman A E 1985 *Lasers* (University Science Books, Mill Valley, California)
[40] Walther A, Amari A, Kröll S and Kalachev A 2009)( Correction: the expression for the group velocity needs to be corrected to $v_g = \frac{2\pi\Gamma}{\alpha}$ in subsection "IVB Slow light effects") *Physical Review A* **80** 012317
[41] Zhang J, Hernandez G and Zhu Y 2008 *Optics Letters* **33** 46–48
[42] Wu H and Xiao M 2007 *Optics Letters* **32** 3122–3124
[43] Sabooni M et al 2013 *manuscript in preparation*.
[44] Shakhmuratov R N, Rebane A, Megret P and Odeurs J 2005 *Physical Review A* **71** 053811
[45] Timoney N, Baumgart I, Johanning M, Varon A F, Plenio M B, Retzker A and Wunderlich C 2011 *Nature* **476** 185–U83